\documentclass[a4paper, fontsize=12pt]{article}

\renewenvironment{abstract}
               {\list{}{\rightmargin\leftmargin}%
                \item[\hspace*{1cm}\small\textbf{Abstract ---}]\relax}
               {\endlist}

\usepackage[top=1in, bottom=1in, left=1in, right=1in]{geometry}
\usepackage[fleqn]{amsmath}
\usepackage{amssymb}
\usepackage{enumerate}
\usepackage{amsthm}
\usepackage[pdftex]{graphicx}
\usepackage{balance}
\usepackage{fancyhdr}
\usepackage{sidecap}
\usepackage{appendix}
\usepackage{float}
\usepackage{scrextend}
\usepackage{changepage}
\usepackage{subfigure}
\usepackage{endnotes}

\let\footnote=\endnote

\newtheorem{Theorem}{Theorem}

\newtheorem{Postulate}[Theorem]{Postulate}

\newtheorem{Corollary}[Theorem]{Corollary}

\begin{document}

\title{\textbf{Berkelian Idealism Regarding Properties in Orthodox Quantum Mechanics, and Implications for Quantum Gravity}}

\author{
        Marcoen J.T.F. Cabbolet\footnote{E-mail: Marcoen.Cabbolet@vub.ac.be}\\
        \small{\textit{Department of Philosophy, Vrije Universiteit Brussel}}\\
        \small{\textit{Pleinlaan 2, 1050 Brussels (Belgium)}}
        }
\date{}

\maketitle

\begin{abstract}\footnotesize Referring to the $18^{\rm th}$ century idealism of George Berkeley in which an object exists if and only if it is observed, this note shows that orthodox quantum mechanics (OQM) entails a Berkelian idealism regarding properties (BIRP): a quantum `has' a property \emph{X} with quantitative value \emph{x} if and only if the property \emph{X} has just been measured with outcome \emph{x}. It is then impossible to recontextualize GR's principle of curvature in any quantum framework that implies this BIRP, for a quantum cannot curve space-time if it doesn't have a definite energy---which is supposed to be the cause of curvature---in absence of observation to begin with. Concluding, it is ruled out that a quantum theory of gravity, in which GR's principle of curvature is built in as a fundamental physical principle, can be developed in any framework implying this BIRP.
\end{abstract}

\noindent The famous dictum \emph{esse est percipi} (``to be is to be perceived''), put forward by Berkeley in his 1710 book \emph{Treatise Concerning the Principles of Human Knowledge}, refers to the idea that an object exists \textbf{if and only if} it is observed. Although this idea is nowadays considered highly implausible---to say the least---the purpose of this short note is (i) to show that a form of Berkelian idealism is entailed by orthodox quantum mechanics (OQM), and (ii) to discuss how that poses an issue for the development of a quantum theory of gravitation.

If we look at the ontology of `quanta', then of course OQM does not deny that a quantum \emph{exists} in absence of observation. The State Postulate, namely, dictates that the state of a quantum is represented by a wave function $\psi$ with norm $\|\psi\|= 1$, and the statistical interpretation of the latter means that OQM predicts \emph{with certainty} that the quantum will turn up somewhere if we look in the whole space. Now by definition, `completeness' of a theory implies that every object predicted with certainty by the theory must have a counterpart in physical reality \cite{EPR}. Ergo, if we view OQM as a complete theory, then we implicitly take the position that a quantum with a wave function $\psi$ for which $\|\psi\|= 1$ has a counterpart in physical reality---meaning that the quantum exists \emph{even if} it is not observed.

To get to the Berkelian Idealism, we must have a look at the postulates of OQM that make statements about properties. Regarding properties that have more than one possible value\footnote{E.g. if we measure the position of an electron, the obtained value is one out of a range of possibilities.}, we then find the following corollary, which has earlier been hinted at by Muller in \cite{Muller0}:

\begin{Corollary}\label{cor:1}
OQM entails a Berkelian idealism regarding properties (BIRP): absent special preparations, a quantum `has' a property $X$ with quantitative value $x_j$ \textbf{if and only if} a measurement of the property $X$ has just been done with outcome $x_j$.
\end{Corollary}

\paragraph{Proof:} This corollary follows from two postulates of OQM, the Standard Property Postulate (SPP) and the Projection Postulate (PP):

\begin{Postulate}{\rm(SPP)} a quantum `has' a property $X$ with quantitative value $x_j$ \textbf{if and only if} it is in the eigenstate $| \ x_j \ \rangle$ of the associated operator $\hat{X}$ {\rm \cite{Muller2}}.\footnote{The use of operators as representations of observable properties has been criticized in \cite{Daumer}; however, this note is about OQM, so with observable properties represented by operators.}$^,$\footnote{The SPP, in particular the ``only if'' part, has been criticized \cite{Muller3}. However, the SPP as formulated is essential to OQM: if we remove the ``only if'' part then we \emph{depart} from the framework of OQM. This note is about OQM, so with the SPP as stated.}
\end{Postulate}

\begin{Postulate}{\rm(PP)} \textbf{if} a measurement of the property $X$ has been done with outcome $x_j$, \textbf{then} immediately after the measurement the quantum is in the eigenstate $| \ x_j \ \rangle$ of the associated operator $\hat{X}$ {\rm\cite{VonNeumann1}}.
\end{Postulate}

\begin{figure}[h!]
\centering
\includegraphics[width=0.80\textwidth]{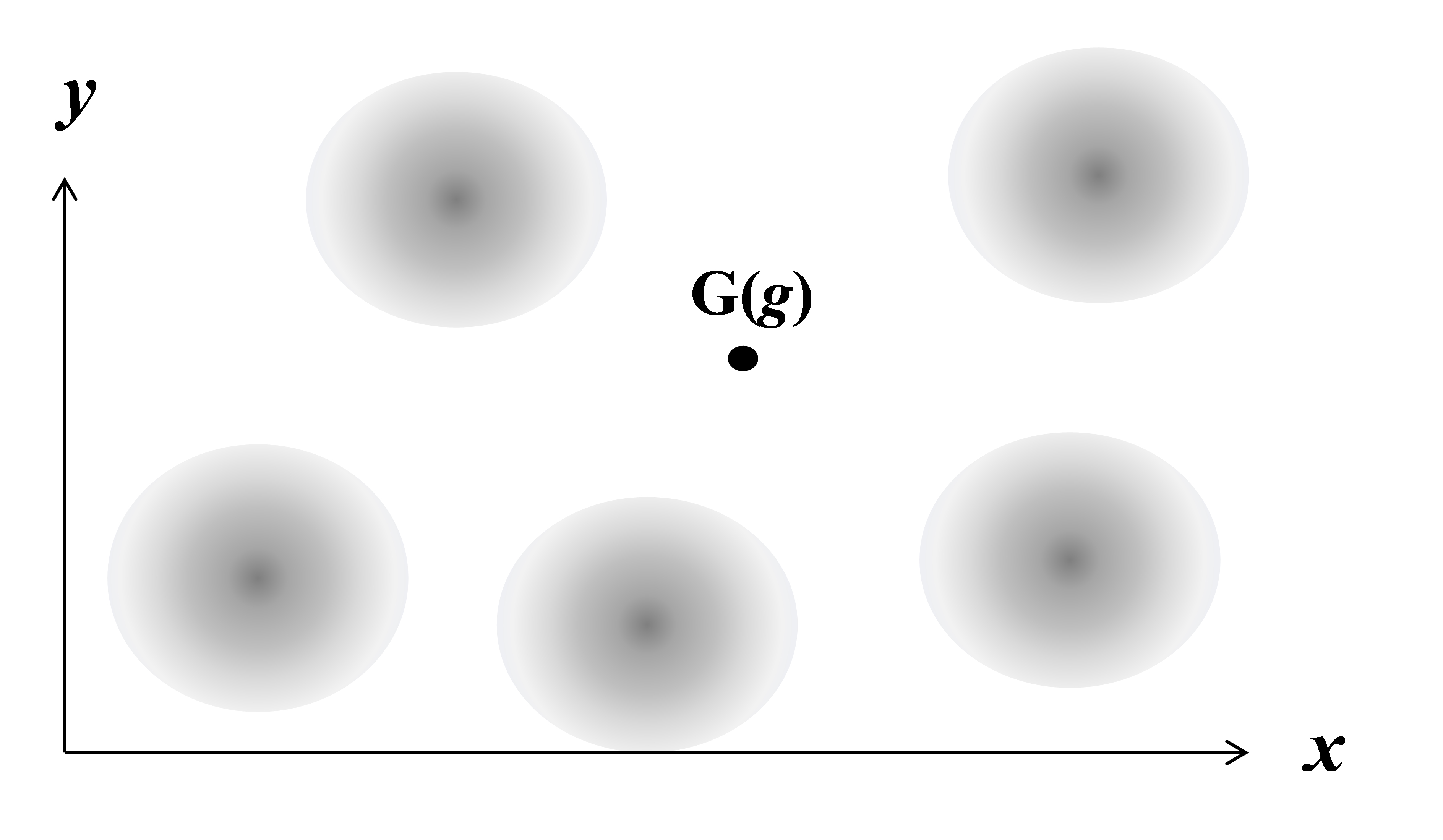}
\caption{\footnotesize \textbf{Illustration of the issue for quantum gravity.} The picture shows a system of five quanta at a given time $t$. Horizontally the spatial $x-$axis, vertically the spatial $y-$axis; the $z-$axis is suppressed. The blurred spots represent the probability distributions of the quanta, with a darker tint indicating a higher probability of being found at that position. So at the spatiotemporal position indicated by the black dot, space-time does have the property curvature with a definite value $G$ regardless whether we make a measurement or not. But without measurement, the quanta do not `have' a definite energy, which is supposed to be the cause of that curvature. Therefore, the principle of curvature cannot be recontextualized in the framework of OQM.}
\label{fig:1}
\end{figure}

\noindent Obviously, the logical form of the SPP is that of a proposition of the type
\begin{equation}\label{eq:1}
P \leftrightarrow E
\end{equation}
with the proposition letters $P$ and $E$ denoting `a quantum has a property $X$ with quantitative value $x_j$' and `the quantum is in the eigenstate $| \ x_j \ \rangle$ of the associated operator $\hat{X}$', respectively. The PP, obviously, has the logical form of an implication
\begin{equation}\label{eq:2}
E \leftarrow M
\end{equation}
with the proposition letter $M$ denoting `a measurement of the property $X$ has just been done with outcome $x_j$'. In addition, the PP is \textbf{the only} postulate of OQM that tells us \emph{how} a quantum can get in the required eigenstate $| \ x_j \ \rangle$: it is, thus, ruled out that the quantum is in the eigenstate $| \ x_j \ \rangle$ of the operator $\hat{X}$ \emph{without} a measurement of the property $X$ having been done with outcome $x_j$.\footnote{Some quantum physicists have asserted that in certain cases a quantum that is in a superposition of eigenstates can get into an eigenstate by temporal evolution: these special cases are excluded by the condition `absent special preparations' in Cor. \ref{cor:1}.} In other words, because the PP is the only postulate of OQM that tells us how a quantum can get in the required eigenstate, we have the additional proposition
\begin{equation}\label{eq:3}
\neg(E \wedge \neg M)
\end{equation}
which is equivalent to
\begin{equation}\label{eq:4}
E \rightarrow M
\end{equation}
Now expressions (\ref{eq:2}) and (\ref{eq:4}) yield
\begin{equation}\label{eq:5}
E \leftrightarrow M
\end{equation}
From expressions (\ref{eq:1}) and (\ref{eq:5}) we get
\begin{equation}\label{eq:6}
P \leftrightarrow M
\end{equation}
This expression (\ref{eq:6}) is precisely the BIRP. Ergo, OQM entails a BIRP. \hfill $\Box$\\
\ \\
So on the one hand, the `only if' part of the BIRP \emph{forbids} to say that a quantum `has' a certain quantitative property when that property hasn't been measured (again, absent special preparations). On the other hand, the `if' part of the BIRP guarantees that the quantum `has' the property $X$ with value $x$ upon a measurement of that proprty $X$ with outcome $x$. This BIRP reflects an essential aspect of the concept of a quantum, that of course remains intact when we make relativistic corrections: these concern the measured \emph{value} of a property, but not the criterion for when a quantum does or doesn't `have' a quantitative property.\\
\ \\
This BIRP poses an issue for the development of a quantum theory of gravitation. An important physical principle in this context is namely the principle of curvature as stated by General Relativity (GR): space-time is curved \textbf{due to} the energy of objects (particles) as expressed by the Einstein field equations. The principle is backed up by empirical evidence, but if we want to develop a quantum theory of gravity and we want to recontextualize this physical principle in the framework of quantum theory, then we stumble on a problem. Namely, the space-time of GR is not a substance, not an object: the fact that it has a metric doesn't make it an object---it is the void between the objects. That being said, it cannot be an individual in the quantum-theoretical ontology that has a wave-function. So we cannot speak of a `metric operator' $\hat{g}$: the metric is not an observable property that a quantum `has' (or can `have'). So on the one hand, space-time is itself not a quantum object that is subjected to this BIRP: therefore, at any spatiotemporal position $(\vec{x}, t)$ space-time has a definite curvature regardless whether it is measured or not\footnote{The fact that space-time has a definite metric on it at every spatiotemporal position doesn't make it an `absolute' space-time. For an observer $\mathcal{O}$, the metric at $(\vec{x}, t)$ is $g$, and for an observer $\mathcal{O}^\prime$, the metric at $(\vec{x}^\prime, t^\prime)$---with $(\vec{x}, t)$ and $(\vec{x}^\prime, t^\prime)$ referring to the same event---is $g^\prime$. But we not necessarily have $g = g^\prime$, so space-time is not absolute.}. But on the other hand, the quanta that populate the universe do not `have' a definite energy (or gravitational mass), which is supposed to be the cause of that curvature, unless the energy is measured: that the BIRP---so no measurement, no cause of curvature. That's the problem. See Fig. \ref{fig:1} for an illustration.

Of course, one can approach the problem purely pragmatically and develop an equation by which the curvature emerges, for example, from expectation values of momenta of the quanta as expressed by this equation:
\begin{equation}
G^{\mu\nu} = T(\langle p \rangle_1, \langle p \rangle_2, \langle p \rangle_3, \ldots)^{\mu\nu}
\end{equation}
This equation is mathematically well-defined, so one might be inclined to think that this solves the problem. However, the expectation values $\langle p \rangle_j$ refer to expected outcomes of experiments: these are not properties that the quanta `have' in absence of measurement. So this ``solution'' would be \textbf{conceptually incoherent}: it is not a solution at all.\\
\ \\
\noindent The main conclusion from the foregoing is that the principle of curvature of GR is fundamentally incompatible with any quantum framework implying the Berkelian idealism regarding properties. This incompatibility comes on top of inconsistencies between GR and quantum theory that have already been identified, see e.g. \cite{Thiemann} for an overview. But the present incompatibility is at the conceptual level, meaning that it cannot be resolved by altering calculational procedures or changing mathematical representations of the concepts: either the view on space-time of GR is false, or the orthodox notion of a quantum is false, or both are false.

Given that the principle of curvature is a relation between things as basic as length, mass, and time, on can take the position that is a \emph{fundamental} physical principle. From that point of view, the present reasoning yields a modus tollens kind of argument against quantum theory: we have seen that the BIRP doesn't work for the principle of curvature, so from a pragmatic point of view the BIRP cannot be true; so if we then have a quantum framework that implies this BIRP, then we are forced to conclude that this framework isn't true either.

On the other hand, if one does want to develop a quantum theory of gravity, then there is no other option to take the position that the world view of GR \emph{emerges} from something more fundamental. But the question is then: what is \emph{more fundamental} than things like mass, length, and time? The only possibility seems to be to take a \emph{substantival} approach to space-time, thus reviving the idea of an aether.

\paragraph{Acknowledgements} This work has been facilitated by the Foundation Liberalitas. Furthermore, the idea to give the name `Berkelian idealism regarding properties' to the if-and-only-if statement in Cor.~\ref{cor:1} emerged from a discussion with F.A. Muller (Erasmus University Rotterdam) about his unpublished work \cite{Muller0}, in which he suggests to describe the relation between OQM and Berkeley's idealism.

\theendnotes

\end{document}